\documentclass[12pt]{article}
\usepackage{amsmath,amssymb}

\def\a{\alpha'}

\begin{document}

\begin{titlepage}
\begin{center}

\vskip 20mm

{\Huge One loop superstring effective actions and $d=4$
supergravity}

\vskip 10mm

Filipe Moura

\vskip 4mm

{\em Security and Quantum Information Group - Instituto de
Telecomunica\c c\~oes\\ Instituto Superior T\'ecnico, Departamento
de Matem\'atica \\Av. Rovisco Pais, 1049-001 Lisboa, Portugal}

\vskip 4mm

{\tt fmoura@math.ist.utl.pt}

\vskip 6mm

\end{center}

\vskip .2in

\begin{center} {\bf Abstract } \end{center}
\begin{quotation}\noindent

We review our recent work on the existence of a new independent
${\cal R}^4$ term, at one loop, in the type IIA and heterotic
effective actions, after reduction to four dimensions, besides the
usual square of the Bel-Robinson tensor. We discuss its
supersymmetrization.
\end{quotation}

\vfill


\end{titlepage}

\eject

\section{${\cal R}^4$ terms in ten dimensions}
\indent

The superstring $\a^3$ effective actions contain two independent
terms $X, Z$ which involve only the fourth power of the Weyl tensor,
given by
\begin{equation}
X:=t_8 t_8 {\cal W}^4, \,\, Z:=-\varepsilon_{10} \varepsilon_{10}
{\cal W}^4. \label{ixiz}
\end{equation}
For the heterotic string two other ${\cal R}^4$ terms $Y_1$ and
$Y_2$ appear at order $\a^3$ \cite{Peeters:2000qj, Tseytlin:1995bi,
deRoo:1992zp}:
\begin{equation}
Y_1 := t_8 \left(\mbox{tr} {\cal W}^2\right)^2, \,\, Y_2 := t_8
\mbox{tr} {\cal W}^4 = \frac{X}{24} + \frac{Y_1}{4}. \label{y1y2}
\end{equation}
Each $t_8$ tensor has eight free spacetime indices, acting in four
two-index antisymmetric tensors as defined in \cite{Gross:1986iv,
Grisaru:1986px}.

The effective action of type IIB theory must be written, because of
its well known SL$(2,{\mathbb Z})$ invariance, as a product of a
single linear combination of order $\a^3$ invariants and an overall
function of $\Omega= C^0 + i e^{-\phi},$ $C^0$ being the axion and
$\phi$ the dilaton. The ${\cal R}^4$ terms of this effective action
are given in the string frame by
\begin{equation}
\left. \frac{1}{\sqrt{-g}} {\mathcal L}_{\mathrm{IIB}}
\right|_{\a^3} = -e^{-2 \phi} \a^3 \frac{\zeta(3)}{3 \times 2^{10}}
\left(X - \frac{1}{8} Z \right) - \a^3 \frac{1}{3 \times 2^{16}
\pi^5} \left(X - \frac{1}{8} Z \right). \label{2bea}
\end{equation}
The corresponding $\a^3$ action of type IIA superstrings has a
relative "-" sign flip in the one loop term
\cite{Antoniadis:1997eg}, because of the different chirality
properties of type IIA and type IIB theories, which reflects on the
relative GSO projection between the left and right movers:
\begin{equation}
\left. \frac{1}{\sqrt{-g}} {\mathcal L}_{\mathrm{IIA}}
\right|_{\a^3} = -e^{-2 \phi} \a^3 \frac{\zeta(3)}{3 \times 2^{10}}
\left(X - \frac{1}{8} Z \right) - \a^3 \frac{1}{3 \times 2^{16}
\pi^5} \left(X + \frac{1}{8} Z \right). \label{2aea}
\end{equation}
Heterotic string theories in $d=10$ have ${\mathcal N}=1$
supersymmetry, which allows corrections already at order $\a$,
including ${\mathcal R}^2$ terms. These corrections come both from
three and four graviton scattering amplitudes and anomaly
cancellation terms (the Green-Schwarz mechanism). Up to order
$\a^3$, the terms from this effective action which involve only the
Weyl tensor are given in the string frame by
\begin{eqnarray}
\left. \frac{1}{\sqrt{-g}} {\mathcal L}_{\mathrm{heterotic}}
\right|_{\a+ \a^3} &=& e^{-2 \phi} \left[\frac{1}{16} \a \mbox{tr}
{\cal R}^2 +\frac{1}{2^9} \a^3 Y_1 - \frac{\zeta(3)}{3 \times
2^{10}} \a^3 \left(X - \frac{1}{8} Z \right) \right] \nonumber
\\ &-& \a^3 \frac{1}{3 \times 2^{14} \pi^5} \left(Y_1+ 4 Y_2
\right). \label{hea}
\end{eqnarray}

Next we will take these terms reduced to four dimensions, in the
Einstein frame, in order to consider them in the context of
supergravity.


\section{${\cal R}^4$ terms in four dimensions}
\indent

In $d=4$, the Weyl tensor can be decomposed in its self-dual and
antiself-dual parts:
\begin{equation}
{\cal W}_{\mu \nu \rho \sigma}= {\cal W}^+_{\mu \nu \rho \sigma} +
{\cal W}^-_{\mu \nu \rho \sigma}, {\cal W}^{\mp}_{\mu \nu \rho
\sigma} :=\frac{1}{2} \left({\cal W}_{\mu \nu \rho \sigma} \pm
\frac{i}{2} \varepsilon_{\mu \nu}^{\ \ \ \lambda \tau} {\cal
W}_{\lambda \tau \rho \sigma} \right). \label{wpm}
\end{equation}
The totally symmetric Bel-Robinson tensor is given by ${\cal
W}^+_{\mu \rho \nu \sigma} {\cal W}^{- \rho \ \sigma}_{\tau \
\lambda}.$ In the van der Warden notation, using spinorial indices,
to ${\cal W}^+_{\mu \rho \nu \sigma},  {\cal W}^-_{\mu \rho \nu
\sigma}$ correspond the totally symmetric ${\cal W}_{ABCD}, {\cal
W}_{\dot A \dot B \dot C \dot D}$ being given by (in the notation of
\cite{Moura:2002ft})
$${\cal W}_{ABCD}:=-\frac{1}{8} {\cal W}^+_{\mu \nu \rho \sigma}
\sigma^{\mu \nu}_{\underline{AB}} \sigma^{\rho
\sigma}_{\underline{CD}}, \, {\cal W}_{\dot A \dot B \dot C \dot
D}:=-\frac{1}{8} {\cal W}^-_{\mu \nu \rho \sigma} \sigma^{\mu
\nu}_{\underline{\dot A \dot B}} \sigma^{\rho
\sigma}_{\underline{\dot C \dot D}}.$$ The decomposition (\ref{wpm})
is written as $${\cal W}_{A \dot A B \dot B C \dot C D \dot D}= -2
\varepsilon_{\dot A \dot B} \varepsilon_{\dot C \dot D} {\cal
W}_{ABCD}  -2 \varepsilon_{AB} \varepsilon_{CD} {\cal W}_{\dot A
\dot B \dot C \dot D};$$ the Bel-Robinson tensor is simply given by
${\cal W}_{ABCD} {\cal W}_{\dot A \dot B \dot C \dot D}$.

In four dimensions, there are only two independent real scalar
polynomials made from four powers of the Weyl tensor
\cite{Fulling:1992vm}, given by
\begin{eqnarray}
{\cal W}_+^2 {\cal W}_-^2 &=& {\cal W}^{ABCD} {\cal W}_{ABCD} {\cal
W}^{\dot A \dot B \dot C \dot D} {\cal W}_{\dot A \dot B \dot C \dot
D}, \label{r441}\\ {\cal W}_+^4+{\cal W}_-^4 &=& \left({\cal
W}^{ABCD} {\cal W}_{ABCD}\right)^2 + \left({\cal W}^{\dot A \dot B
\dot C \dot D} {\cal W}_{\dot A \dot B \dot C \dot D}\right)^2.
\label{r442}
\end{eqnarray}
In particular, the terms $X, Z, Y_1, Y_2,$ when computed directly in
$d=4$ (i.e. expanded only in terms of the Weyl tensor and
restricting the sums over contracted indices to four dimensions),
should be expressed in terms of them. The details of the calculation
can be seen in \cite{Moura:2007ks}: $X - \frac{1}{8} Z$ is the only
combination of $X$ and $Z$ which in $d=4$ does not contain
(\ref{r442}), i.e. which contains only the square of the
Bel-Robinson tensor (\ref{r441}); $Y_1$ (but not $Y_2$) is also only
expressed in terms of (\ref{r441}). We then write the effective
actions (\ref{2bea}), (\ref{2aea}), (\ref{hea}) in four dimensions,
in the Einstein frame (considering only terms which are simply
powers of the Weyl tensor, without any other fields except their
couplings to the dilaton, and introducing the $d=4$ gravitational
coupling constant $\kappa$):
\begin{eqnarray}
\left. \frac{\kappa^2}{\sqrt{-g}} {\mathcal L}_{\mathrm{IIB}}
\right|_{{\cal R}^4} &=& - \frac{\zeta(3)}{32} e^{-6 \phi} \a^3
{\cal W}_+^2 {\cal W}_-^2 - \frac{1}{2^{11} \pi^5} e^{-4 \phi}\a^3
{\cal W}_+^2 {\cal W}_-^2, \label{2bea4} \\ \left.
\frac{\kappa^2}{\sqrt{-g}} {\mathcal L}_{\mathrm{IIA}}
\right|_{{\cal R}^4} &=& - \frac{\zeta(3)}{32} e^{-6 \phi} \a^3
{\cal W}_+^2 {\cal W}_-^2 \nonumber \\ &-& \frac{1}{2^{12} \pi^5}
e^{-4 \phi}\a^3 \left[\left({\cal W}_+^4 + {\cal W}_-^4 \right) +224
{\cal W}_+^2 {\cal W}_-^2 \right], \label{2aea4}
\\ \left. \frac{\kappa^2}{\sqrt{-g}} {\mathcal L}_{\mathrm{het}}
\right|_{{\cal R}^2 + {\cal R}^4} &=& -\frac{1}{16} e^{-2 \phi} \a
\left({\cal W}_+^2 + {\cal W}_-^2 \right) +\frac{1}{64} \left(1-2
\zeta(3)
\right) e^{-6 \phi} \a^3 {\cal W}_+^2 {\cal W}_-^2 \nonumber \\
&-& \frac{1}{3\times2^{12} \pi^5} e^{-4 \phi}\a^3 \left[\left({\cal
W}_+^4 + {\cal W}_-^4 \right) +20 {\cal W}_+^2 {\cal W}_-^2 \right].
\label{hea4}
\end{eqnarray}
These are only the moduli-independent ${\cal R}^4$ terms. Strictly
speaking not even these terms are moduli-independent, since they are
all multiplied by the volume of the compactification manifold, a
factor we omitted for simplicity. But they are always present, no
matter which compactification is taken. The complete action, for
every different compactification manifold, includes many other
moduli-dependent terms which we do not consider here: we are mostly
interested in a ${\mathbb T}^6$ compactification.


\section{${\cal R}^4$ terms and four-dimensional supergravity}
\indent

We are interested in the full supersymmetric completion of ${\cal
R}^4$ terms in $d=4.$ In general each superinvariant consists of a
leading bosonic term and its supersymmetric completion, given by a
series of terms with fermions.

The supersymmetrization of the square of the Bel-Robinson tensor
${\cal W}_+^2 {\cal W}_-^2$ has been known for a long time, in
simple \cite{Deser:1977nt, Moura:2001xx} and extended
\cite{Deser:1978br,Moura:2002ip} four dimensional supergravity. For
the term ${\cal W}_+^4 + {\cal W}_-^4$, which appears at one string
loop in the type IIA and heterotic effective actions (\ref{2aea4})
and (\ref{hea4}), there is a "no-go theorem", which goes as follows
\cite{Christensen:1979qj}: for a polynomial $I({\cal W})$ of the
Weyl tensor to be supersymmetrizable, each one of its terms must
contain equal powers of ${\cal W}^+_{\mu \nu \rho \sigma}$ and
${\cal W}^-_{\mu \nu \rho \sigma}$. The whole polynomial must then
vanish when either ${\cal W}^+_{\mu \nu \rho \sigma}$ or ${\cal
W}^-_{\mu \nu \rho \sigma}$ do.

The derivation of this result is based on ${\mathcal N}=1$ chirality
arguments, which require equal powers of the different chiralities
of the gravitino in each term of a superinvariant. The rest follows
from the supersymmetric completion. That is why the only exception
to this result is ${\cal W}_+^2 + {\cal W}_-^2$, which appears in
(\ref{hea4}): this term is part of the $d=4$ Gauss-Bonnet
topological invariant (it can be made equal to it with suitable
field redefinitions). This term plays no role in the dynamics and it
is automatically supersymmetric; its supersymmetric completion is 0
and therefore does not involve the gravitino.

The derivation of \cite{Christensen:1979qj} has been obtained using
${\mathcal N}=1$ supergravity, whose supersymmetry algebra is a
subalgebra of ${\mathcal N}>1$. Therefore, it should remain valid
for extended supergravity too. But one must keep in mind the
assumptions which were made, namely the preservation by the
supersymmetry transformations of $R$-symmetry which, for ${\mathcal
N}=1$, corresponds to U(1) and is equivalent to chirality. In
extended supergravity theories $R-$symmetry is a global internal
$\mbox{U}\left({\mathcal N}\right)$ symmetry, which generalizes (and
contains) U(1) from ${\mathcal N}=1$.

Preservation of chirality is true for pure ${\mathcal N}=1$
supergravity, but to this theory and to most of the extended
supergravity theories one may add matter couplings and extra terms
which violate U(1) $R$-symmetry and yet can be made supersymmetric,
inducing corrections to the supersymmetry transformation laws which
do not preserve U(1) $R$-symmetry. That was the procedure taken in
\cite{Moura:2007ks}, through the superspace lagrangian
\begin{eqnarray}
{\cal L}&=& \frac{1}{4 \kappa^2} \int \epsilon \left[ \left(
\overline{\nabla}^2 +\frac{1}{3} \overline{R} \right) \left(
\Omega\left(\Phi, \overline{\Phi} \right) + \a^3 \left(b \Phi
\left(\nabla^2 W^2\right)^2 + \overline{b} \overline{\Phi}
\left(\overline{\nabla}^2 \overline{W}^2\right)^2 \right) \right)
\right. \nonumber \\ && \, \, \, \, \, \, \, \, \, \, \, \, \, \, \,
\, \, \, \, \, \, \, \, \, - \left. 8 P\left(\Phi\right) \right]
d^2\theta + \mathrm{h.c.}. \label{r421}
\end{eqnarray}
$\epsilon$ is the chiral density; $\overline{\nabla}^2 +\frac{1}{3}
\overline{R}$ is the chiral projector; $\Phi$ is a chiral
superfield; $$K\left(\Phi, \overline{\Phi}
\right)=-\frac{3}{\kappa^2} \ln \left(-\frac{\Omega\left(\Phi,
\overline{\Phi} \right)}{3} \right), \,\, \Omega\left(\Phi,
\overline{\Phi} \right)=-3+ \Phi \overline{\Phi} + c \Phi +
\overline{c} \overline{\Phi}$$ is a K\"ahler potential and
$$P\left(\Phi\right)=d + a \Phi + \frac{1}{2} m \Phi^2 + \frac{1}{3}
g \Phi^3$$ is a superpotential. $W_{ABC}$ is the chiral ${\mathcal
N}=1$ superfield such that, at the linearized level, $\left.\nabla
_{\underline{D}}W_{\underline{ABC}}\right|={\cal W}_{ABCD} +
\ldots$; ${\cal W}_+^4 + {\cal W}_-^4$ is proportional to $\left.
\left( \nabla^2 W^2 \right)^2 \right| +\mathrm{h.c.}.$ This term
appears in the supersymmetric lagrangian (\ref{r421}) after
eliminating of the auxiliary fields $F=-\frac{1}{2} \left. \nabla^2
\Phi \right|$ and $\overline{F}$.

A similar procedure may be taken in ${\mathcal N}=2$ supergravity,
since there exist ${\mathcal N}=2$ chiral superfields which must be
Lorentz and SU(2) scalars but can have an arbitrary U(1) weight,
allowing for supersymmetric U(1) breaking couplings.

Such a result should be more difficult to achieve for ${\mathcal N}
\geq 3$, because there are no generic chiral multiplets. But for $3
\leq {\mathcal N} \leq 6$ there are still matter multiplets which
one can couple to the Weyl multiplet. Those couplings could
eventually (but not necessarily) break U(1) $R$-symmetry and lead to
the supersymmetrization of (\ref{r442}).

An even more complicated problem is the ${\mathcal N}=8$
supersymmetrization of (\ref{r442}). The reason is the much more
restrictive character of ${\mathcal N}=8$ supergravity, compared to
lower ${\mathcal N}.$ Besides, its multiplet is unique, which means
there are no extra matter couplings one can take in this theory.
Plus, in this case the $R$-symmetry group is SU(8) and not U(8): the
extra U(1) factor, which in ${\mathcal N}=2$ could be identified
with the remnant ${\mathcal N}=1$ $R$-symmetry and, if broken,
eventually turn the supersymmetrization of (\ref{r442}) possible,
does not exist. Apparently there is no way to circunvent in
${\mathcal N}=8$ the result from \cite{Christensen:1979qj}. In order
to supersymmetrize (\ref{r442}) in this case one should then explore
the different possibilities which were not considered in
\cite{Christensen:1979qj}. Since that article only deals with the
term (\ref{r442}) by itself, in \cite{Moura:2007ac} we considered
extra couplings to it and only then tried to supersymmetrize. This
procedure is very natural, taking into account the scalar couplings
that multiply (\ref{r442}) in the actions (\ref{2aea4}),
(\ref{hea4}).

We therefore considered the linearized superfield expressions which,
when expressed in terms of $x-$space fields, would result in
(\ref{r442}), multiplied by some scalar fields. We did not obtain
any expression which was supersymmetric, not even with nonlinear
supersymmetric transformations. Therefore we cannot expect
(\ref{r442}) to emerge from the nonlinear completion of some
(necessarily $\a^3$) linearized superinvariant. One must really
understand the full $\a$-corrections to the Bianchi identities.
Since these corrections are necessarily nonlinear, this means one
cannot supersymmetrize (\ref{r442}) at the linearized level at all.
That never happened for any of the previously known higher-order
terms, which all had its linearized superinvariant.

The main obstruction to this supersymmetrization is that, as we
argued in \cite{Moura:2007ac}, (\ref{r442}) is not compatible with
the full $R-$symmetry group SU(8). Indeed only the local symmetry
group of the moduli space of compactified string theories (for type
II superstrings on ${\mathbb T}^6$, $\mbox{SU}(4) \otimes
\mbox{SU}(4)$) should be preserved by the four dimensional
perturbative string corrections. Most probably, (\ref{r442}) only
has this later symmetry. If that is the case, in order to
supersymmetrize this term besides the supergravity multiplet one
must also consider $U-$duality multiplets, with massive string
states and nonperturbative states. These would be the contributions
we were missing.

But in conventional extended superspace one cannot simply write down
a superinvariant that does not preserve the
$\mbox{SU}\left({\mathcal N}\right)$ $R-$symmetry, which is part of
the structure group. One can only consider higher order corrections
to the Bianchi identities preserving $\mbox{SU}\left({\mathcal
N}\right)$, which would not be able to supersymmetrize (\ref{r442}).
${\mathcal N}=8$ supersymmetrization of this term would then be
impossible.

The fact that one cannot supersymmetrize in ${\mathcal N}=8$ a term
which string theory requires to be supersymmetric, together with the
fact that one needs to consider nonperturbative states (from
$U-$duality multiplets) in order to understand a perturbative
contribution may be seen as indirect evidence that ${\mathcal N}=8$
supergravity is indeed in the swampland, as proposed in
\cite{Green:2007zzb}. This topic deserves further study.

\section*{Acknowledgments} \indent

I thank Pierre Vanhove for very important discussions and
suggestions and the organizers of the conference for the opportunity
to present this work, which has been supported by Funda\c c\~ao para
a Ci\^encia e a Tecnologia through fellowship BPD/14064/2003 and by
FCT and EU FEDER through PTDC, namely via QSec PTDC/EIA/67661/2006
project.

\end{document}